\documentclass[12pt]{article}

\newtheorem{theorem}{Theorem}[section]
\newtheorem{definition}[theorem]{Definition}
\newtheorem{conj}[theorem]{Conjecture}
\def\tr{{\rm Tr}}
\def\proof{\noindent{\bf Proof: }}
\def\RE{{\rm RelEnt}}
\def\riem{{\rm Riem}}
\def\dob{{\rm Dobrushin}}
\def\bur{{\rm Bures}}
\def\sym{{\rm sym}}

\def\geo{{\rm geod}}

\def\half{\frac{1}{2}}
\def\iff{\Longleftrightarrow}
\def\ran{\rm Range}

\def\dtsig{{\mathbf \cdot \sigma}}

\def\bw{{\bf w}}

\def\rmT{{\rm T}}
\def\dbar{\overline{\cal D}}
\def\qed{~~{\bf QED}}
\def\raw{\rightarrow}

\title{Monotone Riemannian Metrics and Relative Entropy
on Non-Commutative Probability Spaces}

\author { Andrew Lesniewski\thanks{partially supported by National Science
Foundation Grant DMS-94-24344 while at the Department of Physics,
Harvard University, Cambridge, Massachusetts.}\\ Paribas Capital Markets \\ 
The Equitable Tower \\ 787 Seventh Avenue \\ New York, NY 10019 USA \\
and \\ Mary Beth Ruskai\thanks{supported by National Science
Foundation Grants DMS-94-08903 and DMS-97-06981.}\\Department of
Mathematics\\ University of Massachusetts Lowell\\Lowell,
MA  01854 USA \\ {\normalsize bruskai@cs.uml.edu} }

\begin{document}

\maketitle

\begin{abstract}
We use the relative modular operator to define a generalized
relative entropy for any convex operator function $g$ on
$(0,\infty)$ satisfying $g(1) = 0$.  We 
show that these convex operator functions can be partitioned
into convex subsets each of which defines a unique
symmetrized relative entropy, a unique family (parameterized
by density matrices) of continuous monotone Riemannian metrics,
a unique geodesic distance on the space of density matrices, and a unique
monotone operator function satisfying certain symmetry and normalization
conditions.  We describe these objects explicitly
in several important special cases, including $g(w) = - \log w$ 
which yields the familiar logarithmic relative entropy.
The relative entropies, Riemannian metrics, and geodesic distances
obtained by our procedure all contract under completely positive,
trace-preserving maps. 
We then define and study the maximal contraction associated
with these quantities.
\end{abstract}

\tableofcontents

\pagebreak

\section{Introduction}

For quantum systems, a state is described by a density matrix $P$, 
i.e., a positive semi-definite operator with trace one. 
We will let $\dbar$ denote the set of {\em density matrices}.
For classical discrete or commutative systems we can identify the
states with the subset of diagonal density  matrices, each of which 
defines a probability vector $p \in {\bf R}^n.$
For commutative systems the usual logarithmic relative entropy
\begin{eqnarray}
H_{\log}(p,q) = \sum_k p_k \log (p_k/q_k)
\end{eqnarray}
can be generalized to
\begin{eqnarray}
H_g(p,q) = \sum_k p_k g(q_k/p_k)
\end{eqnarray}
where $g$ is a convex function on $(0,\infty)$ with $g(1)=0.$
It is well-known that any such $H_g$ contracts under  stochastic
mappings, i.e., 
$H_g(Ap,Aq) \leq H_g(p,q)$ when $A$ is a column stochastic 
matrix.  Cohen, et al, \cite{Cetal} 
defined the entropy contraction coefficient as 
\begin{eqnarray}
  \eta_g(A) = \sup_{p \neq q} \frac{H_g(Ap,Aq)}{H_g(p,q)}.
\end{eqnarray}
In the pair of papers \cite{Cetal,CRS}, it was shown that
for each fixed $A$ all the contraction coefficients associated
with those $g$ which are also {\em operator} convex are
equivalent, more precisely
\begin{theorem}\label{thm:cohen}
If $g$ is operator convex, then
\begin{eqnarray}
\eta_g(A) =  \eta_{\log}(A) = \eta_{(w-1)^2}(A) \leq \eta_{|w-1|}(A).
\end{eqnarray}
\end{theorem}
A summary of these results is given in \cite{Rusk1}.
It suffices to mention here that the observation
\begin{eqnarray}
 \left. \frac{d^2}{dt^2} H_g(p,p+tv) \right|_{t=0} = 
   g^{\prime\prime}(0) \sum_k (v_k)^2/p_k = H_{(w-1)^2}(p,p+v)
\end{eqnarray}
plays a critical role.
The quantity $\sum_k (v_k)^2/p_k$ can also be written
as $M_p(v,v)$ where 
\begin{eqnarray}
   M_p(u,v) = \left. -\frac{\partial^2}{\partial\alpha
\partial\beta}H_g(p+\alpha u,p+\beta v) \right|_{\alpha = \beta = 0}
\end{eqnarray}
is the Riemannian metric corresponding to the Fisher information.
\v{C}encov \cite{Ca,Cv} showed that, for commutative systems, this is the
{\em only} Riemannian metric which satisfies the monotonicity condition
$ M_{Ap}(Av,Av) \leq  M_p(v,v).$
Thus, we can regard Theorem \ref{thm:cohen} as stating that for
operator convex $g$ the maximal contraction 
of the relative entropy and its associated Riemannian metric
are the same.   Since there is only one Riemannian metric, all
the contraction coefficients must be equal.

For quantum systems, the usual logarithmic relative entropy is given by
\begin{eqnarray}
H_{\log}(P,Q) & = & \tr P(\log P - \log Q) \label{def:ent.log}  \\
  & = & \int_0^{\infty} \tr P \left[ \frac{1}{Q+tI} (P-Q) \frac{1}{P+tI} 
      \right] dt  \label{intrep.log}
\end{eqnarray}
with $P,Q$ in ${\cal D}$, the set of 
invertible density matrices.
The integral representation (\ref{intrep.log}) can be used to show that
\begin{eqnarray}
 M_P^{\log}(A,B) & \equiv &
    \left. -\frac{\partial^2}{\partial\alpha \partial\beta}
H_{\log}(P+\alpha A,Q+\beta B) \right|_{\alpha = \beta = 0} \nonumber \\
  & = & \int_0^{\infty} \tr A \left[ \frac{1}{P+tI} B \frac{1}{P+tI} 
  \right] dt . \label{eq:Riemlog}
\end{eqnarray}
Although $M_P^{\log}(A,B)$ is a monotone Riemannian metric, it is not
the only possibility;  $M_P(A,B) = \tr A^* P^{-1}B$ is also 
monotone under completely positive, trace-preserving maps. 
The study of monotone Riemannian metrics on non-commutative
probability spaces was initiated by  Morozova and \v{C}encov \cite{MC}
who did not, however, provide any explicit examples.  A complete
characterization of monotone Riemannian metrics (which includes
the examples above) was given recently by Petz 
 \cite{Pz2,Pz3,PzS1}.  The quantum structure is much richer because
left and right multiplications by $P^{-1}$ are not equivalent.
We will see that $M_P(A,B)$ can always be written in the form
$\tr A^* \Omega_P(B)$ where $\Omega_P$ reduces to multiplication
by $P^{-1}$ when $P$ and $B$ commute.  Thus, for example, 
(\ref{eq:Riemlog}) above gives 
$\Omega_P(B) = \int_0^{\infty} \frac{1}{P+tI} B \frac{1}{P+tI}dt$
which becomes $P^{-1}B$ when $P$ and $B$ commute.

Earlier, Ruskai \cite{Rusk1} tried to extend the entropy contraction
coefficient results of Cohen, et al to non-commutative situations
but obtained only a few preliminary results.
Although one can formally define 
$H_g(P,Q) = \tr P g(Q/P) $ the expression $Q/P$ is ambiguous in the
quantum case.  Using the non-standard definition $Q/P = P^{-1/2}QP^{-1/2}$,
[which yields $H_g(P,Q) = \tr P \log P^{-1/2}QP^{-1/2}$ rather
than (\ref{def:ent.log}) when $g(w) = - \log w$.]
Ruskai and Petz {\cite{PzRu} were able to prove an analogue of
Theorem \ref{thm:cohen} using the fact that
\begin{eqnarray}
 \left. \frac{d^2}{dt^2} H_g(P,P+tA) \right|_{t=0} = 
   g^{\prime\prime}(0) \tr A P^{-1} A
\end{eqnarray}
for all $g$.  In essence, their convention for $Q/P$ always
yields the Riemannian metric $M_P(A,B) = \tr A^* P^{-1} B.$

A better alternative is to use the relative modular operator
introduced by Araki \cite{Ak1,Ak2,BR,OP,Pz1,Pz2} to define $Q/P$.  
This yields the usual logarithmic
entropy (\ref{def:ent.log}) and a rich family of generalized
relative entropies.  Moreover, differentiation then yields the entire
 family of monotone Riemannian metrics found by Petz \cite{Pz2,Pz3,PzS1}.

In this paper we use the relative modular operator to study both
the relative entropies and Riemannian metrics associated with
convex operator functions.  For simplicity, we restrict ourselves
to the matrix algebras associated with finite dimension systems.
Although we do not believe this restriction is essential, it
avoids many technical complications.  [The most serious arises
when the condition $\tr P = 1$ is not compatible with the
requirement that $P$ be invertible (in the sense of having a bounded
inverse in relevant operator algebra).  In that
case, one must restrict the domain of $H_g(P,Q)$ to those pairs
$P,Q$ which have comparable approximate null spaces in some
suitable sense.]   We show that each convex operator function
defines a convex family of relative entropies, a unique
symmetrized relative entropy, a unique family (parameterized
by density matrices) of continuous monotone Riemannian metrics,
a unique geodesic distance on the space of density matrices, and a unique
monotone operator function.  We describe these objects explicitly
in several important special cases, including $g(w) = - \log w.$ 
We then define and study the contraction coefficient associated
with the relative entropy, Riemannian metrics, and metrics.
Finally, we examples showing that these contraction coefficients
can have any value in $[0,1]$ for a suitable stochastic map.

The paper is organized as follows.  In section 2, we give some
basic definitions and results for relative entropy and Riemannian
metrics.  In section 3, we define the corresponding the geodesic
distance, including the Bures metric as a special case.
Finally, in section 4 we study the contraction of all the
quantities under stochastic maps and give bounds on the 
maximal contraction.

\section{Relative Entropy and Riemannian Metrics}

\subsection{Definitions}

We begin by describing the relative modular operator 
 which was originally introduced by Araki to generalize 
the logarithmic relative entropy to type III von Neumann
algebras \cite{Ak1,Ak2,BR,OP,Pz1,Pz2}.  Later, Petz \cite{Pz1}
used it to generalize relative entropy itself.  Let ${\cal D}$
denote the subset of invertible operators in $\dbar.$
Let $P, Q \in {\cal D}$ i.e., $P$ and $Q$ are 
positive definite matrices with $\tr(P)=\tr(Q)=1.$
For matrix algebras, the relative modular operator associated
with the pair of states $\rho_P(A)= \tr(AP)$ and $\rho_Q(A)=\tr(AQ)$
reduces to
\begin{eqnarray}
\Delta_{Q,P}=L_Q R_P^{-1},
\end{eqnarray}
where $L_Q$ and $R_P$ are the left and right multiplication operators,
respectively. Thus $\Delta_{Q,P}(A)=QAP^{-1}$. It is easy to verify
directly that $\Delta_{Q,P}$ is a positive Hermitian operator with respect
to the Hilbert-Schmidt inner product.  

\begin{definition}\label{relgentdef} Let $g$ be an operator convex
function
defined on $(0,\infty)$ such that $g(1)=0$. The relative $g$-entropy of
$P$ and $Q$ is 
\begin{eqnarray}
H_g(P,Q)=\tr(P^{1/2}g(\Delta_{Q,P})P^{1/2}).
\end{eqnarray}
\end{definition}
We will let $\cal G$ denote the set of functions satisfying these
conditions.   Note, however,
 that the argument of $g$, as defined here, is shifted
from that  (which we here denote $g_{\rm C}$) in \cite{Cetal} 
and \cite{CRS} so that  $g_{\rm C}(w) = g(w+1).$
Using standard results from the theory of monotone and convex
operator functions, one can show that $\cal G$ is the class of
functions which can be written in the form
form\begin{eqnarray}\label{convfctrep}
g(w)= a(w-1) + b(w-1)^2 + c\frac{(w-1)^2}{w} +
         \int_0^\infty\frac{(w-1)^2}{w+s}d\nu(s),
\end{eqnarray}
where $b,c >0$ and  $\nu$ is a positive measure on $(0,\infty)$
with finite mass $\int_0^\infty d\nu(s)$.  The term $\frac{(w-1)^2}{w}$
may seem unfamiliar, as it is usually included implicitly in the
integral.  However, writing it separately will be convenient later
and is necessary to ensure that the measure has finite mass.
The function $g(w) = - \log w$ yields the usual logarithmic relative
 entropy (\ref{def:ent.log}) which we continue to denote $H_{\log}(P,Q)$.
The function $g(w) = (w-1)^2$ yields 
\begin{eqnarray}\label{def:quadent}
H_{(w-1)^2} = \tr (P-Q) P^{-1} (P-Q)
\end{eqnarray}
 which we call the
``quadratic relative entropy'';  it plays an extremely important
role in our development.  The function $g(w) = (w-1)^2/(w+1)$ yields
the equally important, but less familiar 
$H_{\bur}(P,Q) = \tr  (P-Q)[L_Q + R_P]^{-1}(P-Q)$, where we use the
subscript  Bures  because (as will be explained in 
section \ref{sect:geo}) it eventually leads
to a geodesic on $\cal D$ referred to as the ``metric of Bures''.

We will study the properties of relative entropy and related
quantities under a class of maps referred to as ``stochastic''.
\begin{definition}\label{stochmap}
A stochastic map  $\phi: {\cal A}_1 \rightarrow  {\cal A}_2$ 
is a completely positive, trace-preserving
map from one von Neumann algebra to another.
\end{definition}
For commutative systems, a stochastic map always corresponds to a
column stochastic matrix, as discussed in the Introduction and in
\cite{Cetal,CRS,Rusk1}.   For non-commutative systems, 
 a partial trace (see section \ref{subsect:phi.examp} or , e.g., 
\cite{LbRu2,Lind}) is an example of  a stochastic map.  General
conditions can be obtained from the Stinespring representation
\cite{St} for completely positive maps or the subsequent work of
Choi \cite{Choi} and Kraus \cite{Kraus} who showed that 
$\phi$ is a completely positive if and only if there exist
operators $\{V_k\}$ with 
$V_k : {\cal A}_1 \rightarrow  {\cal A}_2$ such that 
\begin{eqnarray}\label{ChoiKr.rep}
  \phi(A) = \sum_{k=1}^N V_k A V_k^*.
\end{eqnarray}
The condition that $\phi$ is trace preserving is then 
$\sum_k V_k^* V_k = I$ [and {\em not} $\sum_k V_k V_k^* = I,$
which is the condition that $\phi$ is unital, i.e., $\phi(I_1) = I_2.$]
For algebras with trace (as is the case here) one can use the
Hilbert-Schmidt inner product 
$\langle A, B \rangle = \tr A^* B$ to define the adjoint
$\widehat{\phi}$ of any completely positive map so that
$\tr A^* \phi(B) = \tr \widehat{\phi}(A)^* B.$  It is then
easy to see that  $\widehat{\phi}(A) = \sum_k V_k^* A V_k$
and that $\phi$ is trace -preserving if and only if
$\widehat{\phi}$ is unital.

\subsection{Relative Entropy}

We begin by defining a relative entropy distance as a bilinear
function on $\cal D $ with the properties we expect of
the relative g-entropy $H_g(P,Q).$  It is sometimes convenient to 
extend our definition from ${\cal D} \times {\cal D}$  to the
somewhat larger set of pairs $P,Q$ of positive definite matrices 
with $\tr P = \tr Q.$

\begin{definition}\label{relentdef}
By a relative entropy {\em distance} 
 we mean a function $H(P,Q)$ satisfying:
\begin{itemize}
\item [a)]  $H(P,Q) \geq 0$ with $H(P,Q) = 0 \Leftrightarrow P = Q$.
\item [b)]  $H(\lambda P, \lambda Q) = \lambda H(P,Q)$ for $\lambda > 0$.
\item [c)]   $H(P,Q) $ is jointly convex in $P$ and $Q$.
\end{itemize}
In addition, we say that the relative entropy is {\em monotone} if
\begin{itemize}
\item [d)]   $H(P,Q) $ decreases under stochastic maps $\phi$,
\end{itemize}
that it is {\em symmetric} if
\begin{itemize}
\item [e)]   $H(P,Q) = H(Q,P)$
\end{itemize}
and that it is {\em differentiable} if
\begin{itemize}
\item [f)]   the function $g(x,y) = H(P+xA,Q+yB)$ is differentiable.
\end{itemize} 
\end{definition}
Conditions (b), (c), and (d) are not independent.  It is well-known that by 
embedding ${\bf C}^{n \times n}$ in 
${\bf C}^{n \times n}\otimes {\bf C}^{2 \times 2}$
and choosing $\phi$ to correspond to the partial trace over 
${\bf C}^2$, one can show that (d) implies the subadditivity relation
\begin{eqnarray}
H(P_1 + P_2, Q_1 + Q_2) \leq H(P_1,Q_1) + H(P_2,Q_2).
\end{eqnarray}
  But for functions satisfying the homogeneity condition (b)
this is equivalent to joint convexity.
Because any stochastic map can represented as a 
partial trace \cite{Lind}, it follows that when (a) and (b) hold,
then (c) $\iff$ (d).  Nevertheless, the properties
of convexity and monotonicity are each of sufficient importance to
justify explicitly stating them separately.

A relative entropy distance (even if symmetric) is not a metric in 
the usual sense, because it need not satisfy the triangle inequality.  
Nevertheless, such quantities have been widely used 
\cite{CZ,JB,VP} to measure the difference
between $P$ and $Q$.  Later, we shall show that every
relative g-entropy defines a relative entropy distance 
which then defines a Riemannian metric and an associated
geodesic distance.
 
\begin{theorem} Every relative g-entropy of the form given in
Definition \ref{relgentdef} is a differentiable monotone relative 
entropy distance in the sense of Definition \ref{relentdef}.
\end{theorem}

\proof Properties (a), (b) and (f) are straightforward; 
(d) is due to \cite{Pz1} and implies (c) by the above remarks.
A simple new proof of (d) is given in Section \ref{subsect:monopf}

\begin{theorem}\label{thm:entpyintrep} For each operator convex function 
$g \in \cal G$, 
\begin{eqnarray}
  H_g(P,Q) & = & \tr (Q-P) \left[ b_g P^{-1} + c_g Q^{-1} \right] (Q-P) 
     \label{eq:entpyintrep}\\  
   &~& ~~~~\int_0^\infty\tr\left((Q-P)\frac{1}{L_Q+sR_P}(Q-P)\right) 
         d\nu_g(s) \nonumber \\
  & = & \tr \left[ (Q-P) R_P^{-1} g(\Delta_{QP}) (Q-P) \right] 
    \label{eq:entformsimp}
\end{eqnarray}
where $b_g, c_g$ and $\nu_g$ are as in (\ref{convfctrep}).
\end{theorem}
\proof
We first observe that
\begin{eqnarray}\label{deltaaction}
(\Delta_{Q,P}-I)(P^{1/2})=(Q-P)P^{-1/2}=R_{P^{-1/2}}(Q-P),
\end{eqnarray}
so that
\begin{eqnarray}
H_{w-1}(P,Q)=\tr \left[ P^{1/2}(Q-P)P^{-1/2}\right]=0,
\end{eqnarray}
and the linear term in (\ref{convfctrep}) does not contribute. 
We also find using (\ref{deltaaction}) again
\begin{eqnarray}
  H_g(P,Q) & = & \langle(\Delta_{Q,P}-I)(P^{1/2}),(\Delta_{Q,P}+sI)^{-1}
    (\Delta_{Q,P}-I)(P^{1/2})\rangle \nonumber \\ 
& = & \tr \left[ (Q-P) (\Delta_{Q,P}+sI)^{-1} R_{P^{-1}}(Q-P) \right] 
    \nonumber \\
& = &  \tr (Q-P) \frac{1}{L_Q+sR_P} (Q-P) .
\end{eqnarray}
Letting $s = 0$ yields
\begin{eqnarray} 
H_{(w-1)^2/w}(P,Q) =\tr\left[(Q-P)Q^{-1}(Q-P)\right] = H_{(w-1)^2}(Q,P)
\end{eqnarray} 
and one easily verifies that 
\begin{eqnarray}
H_{(w-1)^2}(P,Q)=\tr((Q-P)P^{-1}(Q-P)).
\end{eqnarray}
Using these results in (\ref{convfctrep}) gives the desired result
 (\ref{eq:entpyintrep}).

 It is worth pointing out that the cyclicity of the trace
implies that
\begin{eqnarray}
  \tr (Q-P) \frac{1}{R_P + s L_Q} (Q-P) = \tr (Q-P) \frac{1}{L_P + s R_Q} (Q-P), 
\end{eqnarray}
although
\begin{eqnarray*}
  \tr (Q-P) \frac{1}{R_P + s L_Q} (Q-P) \neq \tr (Q-P) \frac{1}{R_Q + s L_P} (Q-P), 
\end{eqnarray*}
in general.
 
One can also use the heat kernel representation
\begin{eqnarray}\label{heatker}
(\Delta_{Q,P}+sI)^{-1}=\int_0^\infty e^{-u(\Delta_{Q,P}+sI)}du,
\end{eqnarray}
to obtain another integral representation of $H_g(P,Q).$
\begin{theorem}\label{thm:heatkerrep}
  Let  $m_g(u)=\int_0^\infty e^{-us}d\nu(s)$ 
denote the Laplace transform of the measure $\nu_g.$  Then
\begin{eqnarray*}
\lefteqn{H_g(P,Q) = b_gH_{(w-1)^2}(P,Q)+ c_g H_{(w-1)^2}(Q,P) +}
~~~~~~~~~~~~~&~&  \\ &~& ~~~~~~~~~~~~~~~~+
  \int_0^\infty H_{(w-1)^2 e^{-uw}}(P,Q)m_g(u)du. 
\end{eqnarray*}
where we formally  extend our definition of $H_g(P,Q)$
to the non-convex function $g(w) = (w-1)^2e^{-uw}.$ 
\end{theorem}
\proof  We use (\ref{heatker}) in (\ref{convfctrep}).
\begin{eqnarray*}
\lefteqn{ \int_0^\infty\langle(\Delta_{Q,P}-I)(P^{1/2}),(\Delta_{Q,P}+sI)^{-1}
(\Delta_{Q,P}-I)(P^{1/2})\rangle d\nu_g(s) } &~&
\\ & = & \int_0^\infty\langle(\Delta_{Q,P}-I)(P^{1/2}),e^{-u\Delta_{Q,P}}
(\Delta_{Q,P}-I)(P^{1/2})\rangle m_g(u)du \\
& = & \int_0^\infty\tr((Q-P)(R_{P^{-1}}e^{-u\Delta_{Q,P}})(Q-P)) m_g(u)du \\
& = & \int_0^\infty  H_{(w-1)^2e^{-uw}}(P,Q)m_g(u)du
\end{eqnarray*}
where we have interchanged
the order of integration and then used  (\ref{deltaaction}) again.

\subsection{Monotone Riemannian metrics}

We now consider the relation between relative g-entropy and
Riemannian metrics.  Note that the set of density matrices
$\cal D$ has a natural
structure as a smooth manifold, so that we can define a
Riemannian metric on its tangent bundle $T_* \cal D$,
whose fibers consist of traceless, self-adjoint matrices or
\begin{eqnarray}\label{tangbund}
  T_P {\cal D} = \{ A = A^* : \tr A = 0 \}.
\end{eqnarray}
\begin{definition}\label{def:Riemmetric}
By a Riemannian metric on $\cal D$, we mean a positive
definite bilinear form $M_P(A,B)$ on $T_P \cal D$ such that
the map  $P \rightarrow M_P(A,A)$ is smooth for each fixed 
$A \in  T_* {\cal D}$.
The metric is {\em monotone} if it contracts under stochastic
maps in the sense
\begin{eqnarray}\label{contract}
 M_{\phi(P)}[\phi(A),\phi(B)] \leq M_P(A,B)
\end{eqnarray}
when $\phi$ is a stochastic map. 
\end{definition}
Note that this definition of monotone requires that the stochastic
map $\phi$ act on the base point (i.e., the indexing density matrix
$P$) as well as the arguments of the bilinear form.
\begin{theorem}\label{thm:riem.metric}  
For each $g \in \cal G$ and density matrix $P \in \cal D$,
\begin{eqnarray}
  M_P^g(A,B) & = &
  \left. -\frac{\partial^2}{\partial\alpha
\partial\beta}H_g(P+\alpha A,P+\beta B) \right|_{\alpha = \beta = 0}
            \label{eq:reimgmetric} \label{eq:Riemdiff} \\
   & = & \langle A,\Omega^g_P(B)\rangle = \tr A ^g_P(B) 
  \label{eq:omegadef}
\end{eqnarray}
defines a Riemannian metric on $T_P {\cal D}$, and a 
positive linear operator  $\Omega^g_P$ on $T_P {\cal D}.$
\end{theorem}

The theorem follows easily from the fact that $R_P$, $L_P$ and
their inverses are positive semi-definite operators with respect
to the Hilbert-Schmidt inner product, e.g., $\tr A^* R_P A > 0$,
and the integral representation
in Theorem \ref{thm:entpyintrep}.  We find
\begin{eqnarray} 
\langle A,\Omega^g_P(B)\rangle 
 & = & (b_g+c_g) \tr [A L_{P}^{-1}(B) + B L_{P}^{-1}(A)] + \nonumber \\
  &~&~+ \int_0^\infty \tr \left[ A (L_P + sR_P)^{-1}(B) + B (L_P + sR_P)^{-1}(A)
     \right] d\nu_g(s) \nonumber \\
  & = &  (b_g+c_g) \tr A [L_{P}^{-1} + R_{P}^{-1}](B) +   \nonumber \\
  &~&~+ \int_0^\infty 
    \tr A \left[ (L_P + sR_P)^{-1} + (R_P + sL_P)^{-1} \right](B) d\nu_g(s) 
      \nonumber \\
  & = & \int_0^\infty \tr A \left[
     (L_P + sR_P)^{-1} + (R_P + sL_P)^{-1} \right](B) N_g(s) ds  \nonumber\\
 & = & \left\langle A ,
   \int_0^\infty\left [ (L_P + sR_P)^{-1} + (R_P + sL_P)^{-1} \right](B)~
    N_g(s) ds  \right\rangle \label{intrepRiem}
\end{eqnarray}
where, for simplicity, we temporarily subsume the quadratic terms
into the integral by defining $N_g$ so that 
  $N_g(s) ds = (b_g+c_g) \delta(s) ds + d\nu_g(s).$  
It is critical that $A$ and $B$ are self-adjoint so that we can
interchange $A$ and $B$ by replacing $L_P$ by $R_P$ as in
\begin{eqnarray}
\tr B L_{P}^{-1}(A) = \tr BP^{-1}A = \tr ABP^{-1} = \tr A R_{P}^{-1}(B).
\end{eqnarray}
This result would not hold if we did not require the perturbations
of $P$ and $Q$ to be self-adjoint.  Given that requirement, the result is
necessarily symmetric in the sense that we get the same result from both
$H_g(P,Q)$ and $H_g(Q,P)$. This is already evident in the quadratic
term, whose coefficient depends only on the sum $b+c$, and will be
discussed further below.

We can now use (\ref{intrepRiem}) to obtain several
explicit formulas for $\Omega_P^g.$
\begin{eqnarray}
\Omega^g_P&=&\int_0^\infty 
\left(\frac{1}{sR_P+L_P}+\frac{1}{sL_P+R_P}\right)
   N_g(s)ds  \label{eq:omega1} \\
&=&\int_0^\infty \frac{1}{sR_P+L_P}\left(N_g(s)+s^{-1}
  N_g(s^{-1})\right)ds\\   \label{eq:omega2}
&=&R_P^{-1}\int_0^\infty \frac{1}{s+\Delta_{P,P}}\sigma_g(s)ds 
   \label{eq:omega3}\\
&=& \int_0^1 \left(\frac{1}{sR_P+L_P}+\frac{1}{sL_P+R_P}\right) 
  \sigma_g(s)ds,  \label{eq:omega4}
\end{eqnarray}
where we have used the change of variable $s \rightarrow s^{-1}$ and
\begin{eqnarray*}\label{sigmadef}
\sigma_g(s)=N_g(s)+s^{-1}N_g(s^{-1}).
\end{eqnarray*}
Note that $\sigma_g(s^{-1})=s\sigma_g(s)$. Then, if we define
\begin{eqnarray}\label{eq:kdefa}
k(\lambda)& = & \int_0^\infty\frac{1}{s+\lambda}\sigma_g(s)ds \\
  & = & \int_0^1 \left[ \frac{1}{s+\lambda} +  \frac{1}{s\lambda + 1}
    \right] \sigma_g(s)ds, \nonumber
\end{eqnarray}
we find that $k(\lambda^{-1})=\lambda k(\lambda)$,  $\Omega^g_P=
R_P^{-1}k(\Delta_{P,P})$, and that $k$ can be expressed in terms of 
$g$ as
\begin{eqnarray}\label{eq:kdefb}
k(w)=\frac{g(w)+wg(w^{-1})}{(w-1)^2}.
\end{eqnarray}  
We will let ${\cal K}$ denote this set of functions, i.e.,
\begin{eqnarray}\label{Kdef}
{\cal K} = \{ k : -k ~\hbox{is operator monotone,}~  
    k(w^{-1}) = w k(w), ~\hbox{and}~ k(1) = 1 \}.
\end{eqnarray}
We have recovered half of Petz's result  \cite{Pz2,Pz3,PzS1} 
that there is a one-to-one correspondence
between symmetric Riemannian metrics and functions of the form
(\ref{eq:kdefa}) which satisfy the normalization condition $k(1) = 1.$
(But note that our $k$ corresponds to $1/f$ in Petz's notation.) 
Our approach also easily yields an explicit expression for both
$\Omega^g_P$ and its inverse.
\begin{theorem}\label{thm:omegainv}
For each $g \in {\cal G}$ and $P \in {\cal D}$, the operator 
$\Omega^g_P$ as defined in Theorem \ref{thm:riem.metric}
satisifes $\Omega^g_P = R_P^{-1}k(L_PR_P^{-1})$ and
$[\Omega^g_P]^{-1} = R_Pf(L_PR_P^{-1})$ where
$k(w)$ is given by (\ref{eq:kdefb}) and $f(w) = 1/k(w).$
\end{theorem}

Although $\Omega^g_P$ is initially defined only on $T_* {\cal D},$
 it can easily be extended to all traceless matrices using the
 natural complexification 
$\tr A = 0 \Longrightarrow A = A_1 + i A_2$ with 
$A_1, A_2 \in T_P {\cal D}$
and then to all of ${\bf C}^{n\times n}$ using linearity and
$\Omega^g_P(I) = P^{-1}I.$  The result is equivalent to using
any of the formulas for $\Omega^g_P$ above together with the
obvious extension of $L_P$ and $R_P$ to all of ${\bf C}^{n\times n}.$
We can summarize this discussion as follows.
\begin{theorem}\label{thm:omega.extend}
For each $g \in {\cal G}$ and $P \in {\cal D}$, the operator 
$\Omega^g_P$ as
defined in Theorem \ref{thm:riem.metric}  can be  extended to a
positive linear operator on ${\bf C}^{n\times n}$ so that
 $M_P^g(A,B) = \tr A^* \Omega^g_P(B)$ defines an inner product
on ${\bf C}^{n\times n}.$  On the other hand, for each $g \in {\cal G}$
and $P \in {\cal D}$ equation (\ref{eq:omega3}) defines a
positive linear operator $\Omega^g_P$ on all of ${\bf C}^{n\times n}$,
 and the bilinear form
$M_P^g(A,B) = \tr A^* \Omega^g_P(B)$
extends to a  monotone Riemannian metric satisfying the symmetry
condition $M_P^g(A,B) = M_P^g(B^*,A^*)$.
\end{theorem}
This result is essentially due to Petz  \cite{Pz2,Pz3,PzS1}, who also 
showed the converse result that every  symmetric monotone Riemannian metric
is of this form.  We give an independent proof of monotonicity
at the end of this section.  That the metric is symmetric
is a consequence of the cyclicity of the trace.

The following result is essentially due to Kubo and Ando \cite{KA} who
developed a theory of operator means.
\begin{theorem}\label{thm:ando}
If $k$ given by (\ref{eq:kdefa}) satisfies $k(1) = 1$, then for all 
$P, Q \in {\cal D}$
\begin{eqnarray}
  R_P^{-1} + L_Q^{-1} \geq   R_P^{-1}k(\Delta_{Q,P}) 
     \geq   \left(R_P+L_Q\right)^{-1}.
\end{eqnarray}
\end{theorem}
\proof  This follows easily from  (\ref{eq:kdefa}),
the  elementary inequality 
\begin{eqnarray}
 \frac{w+1}{2w} \geq \frac{1+t}{2}\left[ \frac{1}{t+w} + \frac{1}{tw + 1}\right]
    \geq \frac{2}{w+1},
\end{eqnarray}
and the fact that the normalization $k(1) = 1$ implies that 
$ 2 \sigma_g(t)/(t+1)$ is a probability measure on $[0,1]$.

As immediate corollaries, we find 
\begin{eqnarray}
& \Omega_P^{(w-1)^2} = L_P^{-1}+ R_P^{-1} \geq  \Omega_P^g 
   \geq \left(R_P+L_P\right)^{-1} = \Omega_p^{\bur}&
     \label{ando.cor.omega}\\
&  M_P^{(w-1)^2}(A,A) \geq  M_P^g(A,A)  \geq  M_P^{\bur}(A,A) &
   \label{ando.cor.riem} \\
&  H_{(w-1)^2}^{\sym}(P,Q)  \geq   H_g^{\sym}(P,Q)  \geq H_{\bur}(P,Q) &
   \label{ando.cor.relent}   \end{eqnarray}
where the superscript indicates the symmetric relative entropy
associated with $g.$ 
Thus $k(w) = 2/(w+1)$ 
corresponds to the minimum symmetric relative entropy and minimum
Riemannian metric among the class studied here.  By contrast, we
will see that $g(w) = (w-1)^2$ corresponds to  $k(w) = (w+1)/(2w)$
so that the quadratic relative entropy is maximal.

The operators $\Omega^g_P$ and $[\Omega^g_P]^{-1}$ are non-commutative
versions of multiplication by $P^{-1}$ and $P$ respectively.
Hence, in view of the cyclicity of the
trace, the following result is not surprising.
\begin{theorem}\label{thm:traceomega}  The operator $\Omega^g_P$
given by (\ref{eq:omega3}) satisfies $\tr~\Omega^g_P(A) = \tr AP^{-1}$
and  $\tr[\Omega^g_P]^{-1}(A) = \tr AP$
\end{theorem}

\proof We first observe that in a basis in which $P$ is diagonal with
eigenvalues $p_k$
\begin{eqnarray}
  [R_P^{-1}\frac{1}{s+\Delta_{P,P}}(A)]_{jk}
    = [\frac{1}{sR_P+L_P}(A)]_{jk}  = \frac{1}{s p_k + p_j} a_{jk}
\end{eqnarray}
so that
\begin{eqnarray}
 [\Omega^g_P(A)]_{jk} = \int_0^{\infty} \frac{a_{jk}}{s p_k + p_j} 
\sigma_g(s) ds .
\end{eqnarray}
Then for every  $g \in {\cal G}, P \in {\cal D}$, and  $A \in T_P{\cal D}$
\begin{eqnarray*}
  \tr~\Omega^g_P(A) & = & \sum_j \int_0^{\infty} \frac{a_{jj}}{s p_j + p_j} 
\sigma_g(s) ds \\
  & = & \sum_j p_j^{-1} a_{jj} \int_0^{\infty} \frac{1}{s+1} \sigma_g(s) ds 
 \\  & = &  k(1)  \tr P^{-1} A =
  \tr P^{-1} A  .
\end{eqnarray*} 
The proof for the inverse is similar.  Since $1/k$ is also operator
monotone, we can use Theorem \label{thm:omegainv} to conclude that
$[\Omega^g_P]^{-1}$ can be written in the form
\begin{eqnarray*}
[\Omega^g_P]^{-1} = a R_P + b L_P -
  \int_0^\infty \frac{R_P^2}{sR_P+L_P} d\mu(s).
\end{eqnarray*}
for some positive measure $\mu$.

\subsection{Correspondence between defining functions}

We now make some remarks on the
relation between $g(w), wg(w^{-1})$, and $k(w)$.  It should be clear
from the development above that every function $g \in \cal G$ 
 defines a Riemannian metric and a function $k$ as in
(\ref{eq:kdefa}) or (\ref{eq:kdefb}).   
If we now consider $\hat{g}(w) = wg(w^{-1})$, it is easy to verify
that $\hat{g}(w) \in \cal G$ as well and that
$H_{\hat{g}}(P,Q) = H_g(Q,P)$.  Thus, the map 
$g(w) \rightarrow  wg(w^{-1})$ has the effect of switching the
arguments of the relative entropy and the function
$g(w) + wg(w^{-1})$ yields the symmetrized relative entropy
$H_g(P,Q) + H_g(Q,P)$.  Now, if we
begin with a function $g$ and relative entropy $H_g(P,Q$, the
differentiation in  (\ref{eq:Riemdiff}) automatically
yields a symmetric result.  Thus, all convex combinations 
$a g(w) + (1-a) \hat{g}(w) $ of $g$ and $\hat{g}(w)$  yield
 the same Riemannian metric and the same function $k \in {\cal K}$.  

Conversely, every $k \in {\cal K}$
defines a unique {\em symmetric} relative entropy
via the function $g^{\sym}(w) = (w-1)^2 k(w)$.  It follows
immediately from the integral representation (\ref{eq:kdefb})
and (\ref{convfctrep}) that  $g^{\sym}(w)$ is also
in $\cal G$
and that $w g^{\sym}(w^{-1}) = g^{\sym}(w)$.  Thus,  $k$ selects from
the convex set of relative entropies associated with a given
$g \in {\cal G}$, the symmetric one.  If we observe that the 
integral representation (\ref{eq:kdefa}) is equivalent to $-k$
being an operator monotone  function, we can summarize the
discussion above as follows.

\begin{theorem}
There is a one-to-one correspondence between each of the following
\begin{itemize}
\item [a)] monotone Riemannian metrics extended to bilinear 
forms via the symmetry condition $M_P^g(A,B) = M_P^g(B^*,A^*)$,
\item [b)]  monotone (decreasing) operator functions satisfying
$k(w^{-1}) = w k(w)$ with the normalization $k(1) = 1$, and 
\item [c)] convex operator functions in $\cal G$  
 which satisfy the symmetry relation $w g(w^{-1}) = g(w).$
\end{itemize}
\end{theorem}

The relations between these are given by (\ref{eq:omega3}),
 (\ref{eq:kdefa}), and  (\ref{eq:kdefb}).
In view of this theorem, it would be appropriate to identify
a given operator $\Omega^g_P$ by using the (unique) symmetric
function $g^{\sym}$.  However, we will continue to use the asymmetric
$g$ for such familiar cases as the logarithm.  One might expect
the one-to-one correspondence to extend to twice-differentiable
symmetric monotone relative entropies.  However, Petz and Ruskai
\cite{PzRu} consider relative entropies of the form
$\tilde{H}_g(P,Q) = \tr P g( P^{-1/2} Q P^{-1/2})$.  This class
of monotone relative entropies can be symmetrized; however, 
differentiation of $H_g$ yields the Riemannian metric 
$M_P^{(w-1)^2}(A,B) = \tr A^* [P^{-1} B+ B P^{-1}]$ 
for all $g \in {\cal G}.$  Thus, in particular
$\tilde{H}_{\log}(P,Q) \linebreak = \tr P \log( P^{-1/2} Q P^{-1/2})$
is an example of a relative entropy distance which is {\em not} a relative
g-entropy in the sense of Definition \ref{relentdef}.
  Another class of distinct relative entropy distances is given by
squares of the geodesic distances introduced in Section 3.
Thus, the properties in Definition 1.3 are not sufficient
to completely characterize the relative g-entropy and
allow us to extend the one-to-one
correspondence in Theorem 2.4 extend to a class of relative entropies.
Although we believe that such an additional condition must
exist, we have not found it.

\subsection{Examples} \label{sect:g.examp}

We now give explicit expressions for the relative entropy,
$\Omega_P^g$ and related quantities in several important
special cases.  Thiese examples will also illustrate the
relation between the functions $g, \hat{g}, g^{\sym}$, and $k$
discussed above.

\noindent{\bf Example 1:}
Take $g(w)=-\log w $. Then $\hat{g}(w) = w \log w$, \linebreak[1]
$g^{\sym} = \linebreak[1]  (w-1)\log w$,  
$k(w)= (w-1)^{-1}\log w$, $N_g(s)=(s+1)^{-2}$ and $\sigma_g(s)=1/(s+1)$.  
Then, $H_{\log}(P,Q)$ is given by (\ref{def:ent.log} ), and
\begin{eqnarray}
H_{\log}^{\sym} & = & H_{(w-1) \log w} = 
   \tr (P-Q) \left[ \log P - \log Q \right] \\
   & = &   \int_0^{\infty} \tr (P-Q) \frac{1}{Q+xI}(P-Q) \frac{1}{P+xI}  dx 
\end{eqnarray} and 
\begin{eqnarray*}
\Omega^{\log}_P&=&
 \int_0^\infty \left(\frac{1}{sR_P+L_P}+\frac{1}{sL_P+R_P}\right)\frac{1}
{(s+1)^2}ds\\
&=& \int_0^\infty\frac{1}{s+1}\frac{1}{L_P+sR_P}ds.
\end{eqnarray*}
Making the change of variables $s\rightarrow sR_P$ in the last integral,
 yields
\begin{eqnarray}
\Omega^{\log}_P=\int_0^\infty\frac{1}{s+L_P}\frac{1}{s+R_P}ds
\end{eqnarray}
so that 
\begin{eqnarray}\label{eq:intrep.log.riem}
\langle A,\Omega^{\log}_P(B)\rangle =
 \tr \int_0^{\infty} A^* \frac{1}{sI+P} B \frac{1}{sI+P} ds,
\end{eqnarray}
a result that we obtained earlier (\ref{eq:Riemlog}) using the
integral representation (\ref{intrep.log}) or
\begin{eqnarray}\label{eq:logintrep}
\log P - \log Q =  \int_0^{\infty}
  \left[\frac{1}{Q+xI} - \frac{1}{P+xI}\right]  dx .
\end{eqnarray}
 In this case, it is also well-known \cite{Lb,OP} that
the inverse operator can be written as
\begin{eqnarray}\label{eq:omegaloginv}
[\Omega^{\log}_P]^{-1}(B) = \int_0^1  P^t B P^{1-t} dt.
\end{eqnarray}

\bigskip
\noindent{\bf Example 2:}
 Take $g(w)=(w-1)^2$.  Then $\widehat{g}(w)=(w-1)^2/w$, \linebreak[1]
$g^{\sym}(w) = \linebreak[1] (w-1)^2(w+1)/w$ and $k(w) =(w+1)/(2w).$
Then $H_{(w-1)^2}(P,Q)$ is given by (\ref{def:quadent}), 
\begin{eqnarray*}
  H_{(w-1)^2}^{\sym}(P,Q) & = & \tr (Q-P)\left[ P^{-1} + Q^{-1} \right] (Q-P),
 \\   \Omega^{(w-1)^2}_P & = & R_P^{-1} + L_P^{-1}
\end{eqnarray*}
and
\begin{eqnarray*}
\langle A,\Omega^{(w-1)^2}_P (A)\rangle  = H_{(w-1)^2}(P,P+A) =
   \tr A P^{-1}  A .
\end{eqnarray*}
The associated function  is the maximal
function satisfying the prescribed conditions.  The operator
$\Omega^{(w-1)^2}_P (B) =  P^{-1}B + B  P^{-1}$ so that
\begin{eqnarray}\label{omegaquad}
\Omega^{(w-1)^2}_P = R_P^{-1} +  L_P^{-1}= R_P^{-1} [R_P + L_P]  L_P^{-1}.
\end{eqnarray}

\bigskip
\noindent{\bf  Example 3:} For $s_0>0$ take 
$g_{s_0}(w)=(w-1)^2/(w+s_0)$.  Then
$\widehat{g}_{s_0}(w)=(w-1)^2/(1+ws_0),$
$g^{\sym}(w) = (w-1)^2(w+1)(1+s_0)/(1+ws_0)(w+s_0),$
$k(w) = (w+1)(1+s_0)/(1+ws_0)(w+s_0),$ and
$N_g(s) = \delta(s-s_0).$  Thus
\begin{eqnarray}\label{omegas0def}
\Omega_P^{g_{s_0}}& = & \frac{1}{s_0R_P+L_P}+\frac{1}{s_0 L_P+R_P}  \\
   & = & (s_0 + 1) [s_0R_P + L_P]^{-1}[R_P+L_P][R_P + s_0 L_P]^{-1}. \nonumber
\end{eqnarray}
When no confusion will result, it will be convenient to employ a slight 
abuse of notation and write $\Omega_P^{s_0}$ for $\Omega_P^{g_{s_0}}.$
The case $s_0 = 1$ is particularly important; we have already seen
that it yields the minimal $k \in {\cal K}.$  Then
$k(w) = \frac{2}{1+w},$  
$g(w) = g^{\sym}(w)=  \frac{(w-1)^2}{w+2}$ and
$\Omega_P^{g_{s_0=1}} \equiv \Omega_P^{\bur} = [R_P + L_P]^{-1}$,
 The corresponding Riemannian metric is
$\langle A,\Omega_P^{\bur}(B)\rangle  = \tr A^* [R_P + L_P]^{-1}(B)$
and the corresponding relative entropy
\begin{eqnarray}
H_{\bur}(P,Q) =  \tr (Q - P) [R_P + L_Q]^{-1}(Q - P) = \tr QXPX
\end{eqnarray}
where $X = [R_P + L_Q]^{-1}(Q - P)$.  Because of the cyclicity of the
trace, $H_{\bur}(P,Q)$ is already symmetric and 
$[R_Q + L_P]^{-1}$ would have given the same result.

\bigskip
\noindent{\bf Example 4:}
Take $g(w)=1-w^\alpha$.  Then 
$k(w) = \frac{ (1  - w^\alpha)(1 - w^{1-\alpha}}{\alpha(1-\alpha) (1-w)^2}$ 
and $N_g(s)=\frac{\sin\pi s}{\pi} (1+s)^{\alpha-2}.$  Thus
\begin{eqnarray*}
H_{1-w^\alpha}(P,Q) = 1 - \tr Q^{\alpha} P^{1-\alpha}.
\end{eqnarray*}
\begin{eqnarray*}
\Omega_P^g=R_P^{-1}\int_0^\infty\frac{1}{sI+\Delta_{P,P}}
\frac{\sin\pi s}{\pi}(1+s)^{\alpha-2}ds.
\end{eqnarray*}
After the change of variables $s \rightarrow sR_P$ this becomes
\begin{eqnarray}
\Omega^g_P=\frac{\sin\pi s}{\pi}\int_0^\infty\frac{1}{L_P+s}
\frac{R_P^{1-\alpha}+s^{1-\alpha}}{(R_P+s)^{2-\alpha}}ds.
\end{eqnarray}

\subsection{Monotonicity proof}\label{subsect:monopf}
We now present a new proof of the monotonicity of the relative entropies
and Riemmanian metrics associated with convex operator functions.
\begin{theorem}\label{thm:monotone}
For every convex operator function $g$ of the type considered here,
both the relative entropy $H_g(P,Q)$ and the corresponding
Riemannian metric are monotone, i.e.
\begin{eqnarray}
  H_g(P,Q) \leq H_g[\phi(P),\phi(Q)], \\
  \langle A \Omega_P^g A \rangle \leq  
    \langle \phi(A) \Omega_{\phi(P)}^g \phi(A) \rangle.
\end{eqnarray}
\end{theorem}
This result is essentially due to Petz \cite{Pz1}.  We give an independent
proof as an immediate corollary of the following theorem and
the integral representations (\ref{eq:entpyintrep}) and (\ref{eq:omega3}).
\begin{theorem}\label{thm:monoineq} If $\phi$ is stochastic
\begin{eqnarray}\label{monoineq} 
 \tr A^* \frac{1}{R_P + s L_Q}A & = &
       \tr \phi\left( A^* \frac{1}{R_P + s L_Q}A \right) \nonumber \\
  & \geq & \tr \phi(A^*) \frac{1}{R_{\phi(P)} + s L_{\phi(Q)}}  \phi(A).
\end{eqnarray}
\end{theorem}
\proof
If $P > 0$, then $\tr A^*PA \geq 0$ and $ \tr A^*AP \geq 0$ so that both
$L_P$ and $R_P$ are positive as operators on the Hilbert-Schmidt space. 
Thus for  $Q> 0$, the operator $R_P + s L_Q$ is also positive. 
Let $X = [R_P + s L_Q]^{-1/2} (A) - [R_P + s L_Q]^{1/2}\widehat{\phi}(B)$
 with $B = [R_{\phi(P)} + s L_{\phi(Q)}]^{-1}  \phi(A)$.  
Then $\tr X^*X \geq 0$ so that
\begin{eqnarray}\label{schwarzpf}
   \tr A^* \frac{1}{R_P + s L_Q}A 
    - \tr A^* \widehat{\phi}(B) - \tr \widehat{\phi}(B^*) A \\
    + \tr \widehat{\phi}(B^*)  [R_P + s L_Q] \widehat{\phi}(B)  \geq 0.
   \nonumber
\end{eqnarray}
Since it is easy to see that
\begin{eqnarray*}
  - \tr A^* \widehat{\phi}(B) - \tr \widehat{\phi}(B^*) A
    =  -2 \tr \phi(A^*) \frac{1}{R_{\phi(P)} + s L_{\phi(Q)}} \phi(A),
\end{eqnarray*}
the desired result will follow if we can show that the last term
in (\ref{schwarzpf}) is bounded above by the right side of (\ref{monoineq}).
We find
\begin{eqnarray*}
  \tr \widehat{\phi}(B^*)  [R_P + s L_Q] \widehat{\phi}(B) 
  & = & \tr \widehat{\phi}(B^*) \widehat{\phi}(B)  P
         +  \widehat{\phi}(B^*) sQ \widehat{\phi}(B)  \\
  & = & \tr \widehat{\phi}(B^*) \widehat{\phi}(B)  P
         + \widehat{\phi}(B) \widehat{\phi}(B^*) sQ  \\
  & \leq &  \tr \widehat{\phi}(B^*B)  P
         + \widehat{\phi}(BB^*) sQ
\end{eqnarray*}
where the inequality follows from positivity of $P$ and $Q$ and
the operator inequality
\begin{eqnarray}
    \widehat{\phi}(B^*) \widehat{\phi}(B) \leq \widehat{\phi}(B^*B),
\end{eqnarray}
which holds for any $B$ because the trace-preserving condition on
$\phi$ gives  $\widehat{\phi}(I_2) = I_1$.  Then using, e.g., 
$\tr \widehat{\phi}(B^*B)  P = \tr B^*B \phi(P)$,  we find
\begin{eqnarray*}
  \tr \widehat{\phi}(B^*)  [R_P + s L_Q] \widehat{\phi}(B) 
   & \leq &  \tr B^*B  \phi(P)  + BB^* s \phi(Q)  \\
& = & \tr B^* [B \phi(P)  + s \phi(Q) B] \\
    & = & \tr B^* [R_{\phi(P)} + s L_{\phi(Q)}] B   = \tr B^* \phi(A) \\
 & = &  \tr \phi(A^*) \frac{1}{R_{\phi(P)} + s L_{\phi(Q)}} \phi(A).
\end{eqnarray*}

It is interesting to observe that the strategy used here is very
similar to that used by Lieb and Ruskai \cite{LbRu2}
to prove a Schwarz inequality for completely positive mappings and,
as a special case, the monotonicity of the quadratic relative
entropy.  At that time, Lieb and Ruskai could use these Schwarz 
inequalities to prove many special cases of the strong subadditivity
of the logarithmic relative entropy, but not the general case.
A complete proof of strong subadditivity \cite{LbRu1} (see also
\cite{Rusk1,W}) seemed to require one of
the convex trace function theorems of Lieb \cite{Lb}.  It is therefore
curious that now, some 25 years later, we have finally found a
way to recover strong subadditivity directly from the Schwarz
strategy of Lieb and Ruskai \cite{LbRu2}.

It should also be noted that Uhlmann had earlier \cite {U1} used
a very different approach (based on interpolation theory) to show the 
logarithmic relative entropy was monotone under a related class 
of mappings that are Schwarz in the sense 
$\phi(A^*A) \geq \phi(A^*) \phi(A)$ and Petz \cite{Pz1}
extended this to other relative entropies.

\section{Geodesic distance}\label{sect:geo}

We now wish to consider the contraction of the relative entropy
and corresponding Riemannian metric under stochastic mappings.
Before doing so, it will be useful to consider the  geodesic distance
 which arises from the Riemannian metrics considered here.

\begin{definition}\label{geodef} Associated with every Riemannian
metric $\langle A,\Omega^g_P(B)\rangle$ of the form (\ref{eq:reimgmetric})
 is a geodesic distance $D_g(P,Q)$  which is defined as
\begin{eqnarray*}
  D_g(P,Q) \equiv 
\inf \int_0^1 \sqrt{\langle \dot S(t),\Omega^g_{S(t)}\dot S(t)\rangle} dt
\end{eqnarray*}
where the infimum is taken over all smooth paths $S(t)$ with
$S(0) = P$ and $S(1) = Q$.
\end{definition}

\begin{theorem} The square $[D_g(P,Q)]^2$ of every geodesic distance of
the form given in
Definition \ref{geodef} is a differentiable monotone relative 
entropy distance in the sense of Definition \ref{relentdef}.
In addition, $D_g(P,Q)$ satisfies the triangle inequality
$D_g(P,R) \leq D_g(P,Q) + D_g(Q,R).$
\end{theorem}

\proof Properties (a), (b) and (e) of Definition \ref{relentdef}
 are readily verified.  
Property (d), i.e,  the monotonicity  $D_g[\phi(P),\phi(Q)[ \leq D_g(P,Q)$ 
can be proven  directly, but also follows easily as a corollary to 
Theorem \ref{supineqthm} below.  The triangle inequality is
standard.  That $D_g(P+xA,Q+yB)$ is differentiable in the sense of 
Definition \ref{relentdef}(f) follows from standard results 
(see, e.g., Theorem 3.6, part (2) of \cite{KN}).  \qed

It is well-known (see, e.g., \cite{UB1,UB2,UB3}) that the metric 
associated with the minimal function $k(w) = \frac{2}{1+w}$ 
discussed in Example 3, is (except for normalization)
 the metric of Bures, i.e.,
$D_{2(w-1)^2/(1+w)}(P,Q) = 4 D^{\bur}(P,Q)$ where
\begin{eqnarray}
 \lefteqn{[D^{\bur}(P,Q)]^2 = \inf 
     \left\{ \tr (W - X)(W-X)^* : WW^* = P, XX^* = Q \right\}}~~~~~~ \nonumber \\
  & = &  2 \left[ 1 - \tr (\sqrt{P}Q\sqrt{P})^{1/2}\right] \\
& \leq & \tr [\sqrt{P} - \sqrt{Q}]^2  = 2 [1 - \tr \sqrt{P}\sqrt{Q}]  
  = H_{1-\sqrt{w}}(P,Q) . 
\end{eqnarray}

It follows immediately from (\ref{ando.cor.omega}) that
\begin{eqnarray}
  D_{(w-1)^2}^{\sym}(P,Q)  \geq  D_g(P,Q)  \geq 4 D^{\bur}(P,Q).
\end{eqnarray}
so that $4D^{\bur}(P,Q)$ gives the minimal geodesic distance of
this type.

\section{Contraction Under Stochastic Maps}

\subsection{Contraction coefficients}

Because the relative entropies, Riemannian metrics, and geodesic
distances all contract under stochastic maps, their maximal
contraction is a well-defined quantity in the following sense.

\begin{definition}
For each fixed convex operator function $g$ of the form given
in Def. \ref{relgentdef} and stochastic map  $\phi$ we define
three entropy contraction coefficients
\begin{eqnarray}
\eta_g^{\RE}(\phi) & = &
    \sup_{P\neq Q}\frac{H_g[\phi(P),\phi(Q)]}{H_g[P,Q]}, \label{etadef.relent}\\
\eta_g^\riem(\phi) & = &  
\sup_P\sup_{A \in T_P {\cal D}}\frac{\langle\phi(A),\Omega^g_{\phi(P)}[\phi(A)]
    \rangle}{\langle A,\Omega^g_P[A]\rangle}, \label{etadef.riem} \\
\eta_g^\geo(\phi) & = & \sup_{P\neq Q} 
     \frac{ [D_g(\phi(P),\phi(Q))]^2 }{[D_g(P,Q)]^2 }. \label{etadef.metric}
\end{eqnarray}
\end{definition}

In \cite{Cetal,CRS} it was shown that for commutative systems,
$\eta_g^\RE(\phi) = \eta_g^\riem(\phi) = \eta_{(w-1)^2}(\phi)$.
Here, we will prove some relations between these  various $\eta$.

\begin{theorem}\label{supineqthm}
The three contraction coefficients defined above satisfy
\begin{eqnarray}
  1 \geq \eta_g^\RE(\phi) \geq \eta_g^\riem(\phi)
     \geq \eta_g^\geo(\phi) .
\end{eqnarray}
\end{theorem}

The intuition behind the second inequality can be seen by 
letting $A = B = Q-P$ in the integral representations of
Theorems 2.2 and 2.3. Then the only
difference between the ratios in (\ref{etadef.relent}) and
(\ref{etadef.riem}) is that the modular operator
in the former is $\Delta_{Q,P}$ while that in the latter is 
$\Delta_{P,P}$.  This would seem to indicate that the first
supremum is taken over a larger set.  However, the two are not
directly comparable because the condition $P \neq Q$ in the
first case precludes the choice  $\Delta_{P,P}$.  Hence,
we consider $Q = P + \epsilon A.$ 

\proof  The upper bound of $1$ follows immediately from Theorem \ref{thm:monotone}.
To prove the second inequality
$\eta_g^\RE(\phi) \geq\eta_g^\riem(\phi)$ we consider, as suggested above,
 $H_g(P,P+\epsilon A) =\tr P^{1/2} g(\Delta_{P,P+\epsilon A})(P^{1/2})$.
Proceeding as in the proof of Theorem \ref{thm:entpyintrep} but with
the shorthand $dN_g(s) = (b_g+c_g) \delta(s) ds + d\nu_g(s)$,  we obtain
\begin{eqnarray*}
H_g(P,P+\epsilon A) & = &  \epsilon^2
  \int_0^\infty\tr\left[ A\frac{1}{L_{P+\epsilon A}+sR_P}(A) \right] 
         dN_g(s) \\
   & = & \epsilon^2
  \int_0^\infty\tr \left[ A\frac{1}{L_P+sR_P}(A) \right] 
         dN_g(s)  + O(\epsilon^3) \\
& = & \epsilon^2\langle \phi(A),\Omega^g_{\phi(P)}(\phi(A))\rangle
  + O(\epsilon^3).
\end{eqnarray*}
Thus
\begin{eqnarray*}
\eta_g^{\RE}(\phi)&=& \sup_{P\neq Q}\frac{H_g[\phi(P),\phi(Q)]}{H_g[P,Q]} \\
&\geq &  \sup_P\sup_{A \in T_* {\cal D}}
     \frac{H_g[\phi(P),\phi(P+\epsilon A)]}{H_g(P,P+\epsilon A)}. 
\end{eqnarray*}
However
\begin{eqnarray*} 
 \frac{H_g[\phi(P),\phi(P+\epsilon A)]}{H_g(P,P+\epsilon A)} =
    \frac{\langle\phi(A),\Omega^g_{\phi(P)}[\phi(A)]\rangle +  O(\epsilon)}
         {\langle A,\Omega^g_P[A]\rangle +  O(\epsilon)}.
\end{eqnarray*}
Since the quantity on the right can be made arbitrary close to 
$\eta_g^{\riem}(\phi)$,
we conclude that $\eta_g^{\RE}(\phi) \geq \eta_g^{\riem}(\phi).$
 Finally, 
to prove the third inequality we first choose $S_o(t)$ to be a
minimizing path for $D_g(P,Q)$, i.e. 
\begin{eqnarray*}
D_g(P,Q) = \int_0^1 \sqrt
   {\langle \dot S_o(t),\Omega^g_{S_o(t)}\dot S_o(t)\rangle} dt .
\end{eqnarray*}
Then, $\phi \circ S_o$ is a smooth path from $\phi(P)$ to  $\phi(Q)$. 
Moreover, the linearity of $\phi$ implies 
that $\frac{d}{dt}\phi \circ S_o(t) = \phi \circ  \dot S_o(t)$.
Thus 
\begin{eqnarray*}
D_g[\phi(Q),\phi(Q)] & \leq & \int_0^1 \sqrt
   {\langle \phi \circ \dot S_o(t),\Omega^g_{\phi \circ S_o(t)} 
       \phi \circ \dot S_o(t) \rangle} dt \\
   & \leq & \ [\eta_g^\riem(\phi)]^{1/2} \int_0^1 \sqrt
  {\langle \dot S_o(t),\Omega^g_{S_o(t)}\dot S_o(t)\rangle} dt  \\
   & = &  [\eta_g^\riem(\phi)]^{1/2}  D_g(P,Q) .
\end{eqnarray*}
Dividing both sides by $D_g(P,Q)$ and taking the supremum of the left hand
side, gives the desired result. \qed

In this case of the first inequality 
$\eta_g^\RE(\phi) \geq \eta_g^\riem(\phi)$, we proved
slightly more, namely, 
that either equality holds or the supremum in $\eta_g^\RE(\phi)$
is actually attained for some non-negative (but not
necessarily strictly positive) density matrices $P, Q$, i.e., strict
inequality implies that there exists  $P \neq Q \in \dbar$
such that 
\begin{eqnarray}\label{eq:etaequal}
H_g[\phi(P),\phi(Q)] = \eta_g^\RE(\phi) H_g(P,Q).
\end{eqnarray}
This follows from the fact that we can always find a maximizing
sequence $ (P_k, Q_k) $ such that 
\begin{eqnarray*}
\lim_{k \rightarrow \infty} \frac{H_g[\phi(P_k),\phi(Q)_k]}{H_g(P_k,Q_k)}
   = \eta_g^{\RE}(\phi) .
\end{eqnarray*}
Since we are in a finite dimensional space, the space of non-negative
density matrices is compact so that we can find a convergent
subsequence $( P_{k_j}, Q_{k_j}) \rightarrow (P, Q).$
Then either $P = Q$ in which case we necessarily have
$\eta_g^\RE(\phi) = \eta_g^\riem(\phi)$ or (\ref{eq:etaequal})
holds.  (Strictly speaking, we must also exclude the possibility
that both $H_g(P_k,Q_k)$ and $H_g[\phi(P_k),\phi(Q_k)]$ diverge
to $\infty$.)  We expect that for most choices of $g$ 
equation (\ref{eq:etaequal}) 
holds only in very special cases [see, e.g., the partial trace
example in Section \ref{subsect:phi.examp} which yield 
$\eta_g^\RE(\phi) = 1 = \eta_g^\riem(\phi).$]  Indeed, even for
commutative  systems, early proofs \cite{AG,Cetal} that equality
holds for $\eta_{\log}(A) = \eta_{(w-1)^2}(A)$ depended on a 
demonstration that (\ref{eq:etaequal}) could not hold in general.

Another special situation occurs for the minimal $g$ which yields the 
Bures metric.  If $P, Q$ commute, then 
\begin{eqnarray*}
  H_{\bur}(P,Q) & \equiv  &
  \tr (P-Q) \big( [L_P+R_Q]^{-1} + [L_Q+R_P]^{-1} \big) (P-Q) \\
  &=& 2 \tr (P-Q) (P+Q)^{-1} (P-Q)  \\ 
 &=& 2 \langle (P-Q),\Omega^{\bur}_{P+Q}[(P-Q)]\rangle .
\end{eqnarray*}
Thus if the supremum for $\eta_{\bur}^\riem(\phi)$ happens
to be attained
for a commuting pair $R,A$ (with $R \in {\cal D}$ and $A \in T_P {\cal D}$)
whose images $\phi(R), \phi(A)$ also commute, then
\begin{eqnarray}\label{eq:bures.attain}
   H_{\bur}[\phi(R+A),\phi(R-A)] = 
      \eta_{\bur}^\riem(\phi) H_{\bur}(R+A,R-A). 
\end{eqnarray}
If $\eta_{\bur}^\RE(\phi) = \eta_{\bur}^\riem(\phi),$
 then this also yields equality in (\ref{eq:etaequal}); however, it does
{\em not} give strict inequality for 
$\eta_{\bur}^\RE(\phi) \geq \eta_{\bur}^\riem(\phi).$  On the
contrary, it seems to offer some heuristic support for equality.

We expect that in those exceptional situation in which
the supremum $\eta_g^\RE(\phi)$ is attained the result is
equal to $\eta_g^\riem(\phi)$ so that equality always holds, 
at least for the first inequality in Theorem \ref{supineqthm}.

Recall that many common choices for $g$ [e.g., $g(w) = (w-1)^2$
or $g(w) = -\log w$] do not yield a symmetric relative entropy,
 i.e., $H_g(P,Q) \neq H_g(Q,P).$  This raises the question
of whether or not the entropy contraction coefficient 
[which we denote 
$\eta_g^{\sym}(\phi) \equiv \eta_{g(w) + w g(w^{-1})}^{\RE}(\phi)$]
for the symmetrized relative entropy 
\begin{eqnarray}
H_g^{\sym}(P,Q) = H_g(P,Q) + H_g(Q,P) = H_{g(w) + w g(w^{-1})}(P,Q)
\end{eqnarray}
is the same as $\eta_g^{\RE}(\phi).$
Although we believe equality holds, we can only prove that
\begin{eqnarray}\label{ineq:sym.eta}
\eta_g^{\sym}(\phi) \leq \eta_g^{\RE}(\phi).
\end{eqnarray}
Nevertheless, Theorem \ref{supineqthm} holds for any $g$.  
In fact, since
there is a unique Riemannian metric associated with all $g$ which
yield the same symmetrized relative entropy, we have
$\eta_g^{\RE}(\phi) \geq \eta_g^{\sym}(\phi) \geq \eta_g^{\riem}(\phi).$  
To prove (\ref{ineq:sym.eta})  it suffices to observe that
\begin{eqnarray*}
H_{g(w) + w g(w^{-1})}(P,Q) = H_g^{\sym}(P,Q) = H_g(P,Q) + H_g(Q,P)
\end{eqnarray*}
so that
\begin{eqnarray*}
H_g^{\sym}[\phi(P),\phi(Q)] & = &   
  H_g[\phi(P),\phi(Q)] +  H_g[\phi(Q),\phi(P)] \\
 &  \leq & \eta_g^{\RE}(\phi) H_g(P,Q) + \eta_g^{\RE}(\phi) H_g(Q,P) \\
   & = &   \eta_g^{\RE}(\phi) H_g^{\sym}(P,Q).
\end{eqnarray*}
In the case of the quadratic entropy, it easily follows that
$\eta_{(w-1)^2}^{\riem}(\phi) = \eta_{(w-1)^2}^{\RE}(\phi)
  = \eta_{(w-1)^2}^{\sym}(\phi).$

Finally, we note that the joint convexity of relative
entropy, Riemannian metrics, and $[D_g(P,Q)]^2$ imply
that the corresponding contraction coefficients are
convex in $\phi.$  (Although we did not explicitly state the
joint convexity for $M_P(A,A)$ it is an easy consequence of
homogeniety and contraction under partial traces.)
\begin{theorem}
For each fixed $g \in {\cal G}$, each of the contraction coefficients
$\eta_g^{\RE}(\phi)$, $\eta_g^{\riem}(\phi)$, and $\eta_g^{\geo(\phi)}$
is convex in $\phi.$ 
\end{theorem}
\proof Since the argument is straightforward, we give details
only for the relative entropy.  Let $\phi = x \phi_1 + (1-x) \phi_2.$
\begin{eqnarray*}
  H_g[\phi(P),\phi(Q)] & = & H_g[x \phi_1(P) + (1-x) \phi_2(P),
                    x \phi_1(Q) + (1-x) \phi_2(Q)] \\
   & \leq & x H_g[\phi_1(P),\phi_1(Q)] + (1-x)  H_g[\phi_2(P),\phi_2(Q)] \\
   & \leq & x~ \eta_g^{\RE}(\phi_1) H_g(P,Q) + (1-x)~ \eta_g^{\RE}(\phi_2)
          H_g(P,Q)  \\
   & = & \left[ x ~\eta_g^{\RE}(\phi_1) + (1-x) ~\eta_g^{\RE}(\phi_2) \right]
          H_g(P,Q) . 
\end{eqnarray*}
Dividing both sides by $H_g(P,Q)$ implies 
\begin{eqnarray*}\eta_g^{\RE}(\phi) \leq x \eta_g^{\RE}(\phi_1) + (1-x)
   \eta_g^{\RE}(\phi_2).  \qed
\end{eqnarray*} 

\subsection{Eigenvalue formulation of $\eta_g^\riem(\phi)$}\label{subs:eval}
We now show how $\eta_g^\riem(\phi)$ is related to
 the following set of eigenvalue problems:
\begin{eqnarray}
\big[\widehat\phi\circ\Omega^g_{\phi(P)}\circ\phi\big](A)=\lambda
\Omega^g_{P}(A).\label{eigenvprob}
\end{eqnarray}
In view of Theorem \ref{thm:omega.extend}, this is a
well-defined linear eigenvalue problem on ${\bf C}^{n \times n}$ for
each fixed pair $\phi$ and $P$. The following remarks are
easily verified.
\begin{itemize} 
\item[a)] The eigenvalue problem (\ref{eigenvprob}) can be
rewritten as $\Phi_P^g \circ \phi(B) = \lambda B$  where
\begin{eqnarray}\label{eq:bigphi}
 \Phi_P^g \equiv (\Omega_P^g)^{-1} \circ \widehat{\phi}\circ\Omega_{\phi(P)}^g.  
\end{eqnarray}
Furthermore, $\Phi_P^g$ is trace-preserving.  This follows from
Theorem \ref{thm:traceomega} and
\begin{eqnarray*}
\tr \Phi_P^g(B) & = & \tr P \widehat{\phi}\circ\Omega_{\phi(P)}^g(B) 
   = \langle  P,  \widehat{\phi}\circ\Omega_{\phi(P)}^g(B) \rangle\\
   & = & \langle \phi(P), \Omega_{\phi(P)}^g(B) \rangle =
     \langle \Omega_{\phi(P)}^g[\phi(P)], B \rangle\\
    & = & \langle I, B \rangle = \tr B
\end{eqnarray*}

\item[b)]  We can assume without loss of generality that matrices which
are eigenvectors in  (\ref{eigenvprob})  are self-adjoint, i.e.,
that $A = A^*.$ Indeed,
it is easy to check that the operator 
$\Omega_P^{g_{s_0}}(A) = (sR_P + L_P)[R_P + L_P]^{-1}(R_P + sL_P)(A)$
satisfies  $[\Omega_P^{g_{s_0}}(A)]^* = \Omega_P^{g_{s_0}}(A^*).$ 
Therefore, the operators  
$\Omega_P^g, \Omega_{\phi(P)}^g, \phi, \widehat{\phi}$
and $\Phi_P^g$ all  map adjoints to adjoints. 

\item[c)] For each fixed $P$, the eigenvalue equation is
satisfied with $A = P$ and eigenvalue $\lambda = 1$ which is the
largest eigenvalue.  The operators on both sides of (\ref{eigenvprob})
are self-adjoint  (in fact, positive definite) with respect to the
Hilbert-Schmidt inner product
and the corresponding orthogonality condition for the other
eigenvectors reduces to  $\tr A = 0.$

\end{itemize}

In view of these observations, it is easy to conclude from the max-min 
principle that the second-largest eigenvalue $\lambda_2^g(\phi, P)$ 
satisifies 
\begin{eqnarray}\label{lambdasup}
\lambda_2^g(\phi,P)=\sup_{A \in T_P {\cal D}}
 \frac{\langle\phi(A),\Omega^g_{\phi(P)}[\phi(A)]
          \rangle}{\langle A,\Omega^g_P[A]\rangle}
\end{eqnarray}
for each fixed $P.$  Then taking the supremum over ${\cal D}$ yields
\begin{theorem}\label{thm:etariemsup}  For each $g \in {\cal G}$ and
stochastic map $\phi$
\begin{eqnarray}\label{eq:etariemsup}
\eta_g^{\riem}(\phi)=\sup_{P \in {\cal D}} \lambda_2^g(\phi, P).
\end{eqnarray}
\end{theorem}

We have already observed that  every $\Omega_P^g$ can be
regarded as a non-commutative
variant of multiplication by $P^{-1}$.  Indeed, if both pairs of
operators $ P,A $ and  $ \phi(P),\phi(A) $ associated with a 
particular eigenvalue  commute for some $g$, then  
$\Omega_P^g(A) = R_{P^{-1}}(A) = L_{P^{-1}}(A)$ 
for {\em all} $g$ and the corresponding eigenvalue equations
are the same.  It may be tempting to conjecture that the eigenvalue
equations for different $g$ are related by a similarity transform,
which would then imply that all $\lambda_2(\phi,P)$ are equal
so that all $\eta_g^{\riem}(\phi)$ are identical.
However, for a given fixed $P$, $R_P$ and $L_P$ commute, which implies that 
 $\Omega_P^g$ and $\Omega_P^h$ commute for any pair of functions
$g$ and $h$.  Since commuting operators are simultaneously
diagonalizable and similar operators have the same eigenvalues,
this would imply that all of the eigenvalue operators
$B \rightarrow \left[ (\Omega_P^g)^{-1}\circ
\widehat{\phi}\circ\Omega_{\phi(P)}^g \circ \phi \right] (B)$
are identical.  This is easily seen to be false in specific examples.
Moreover, as discussed at the end of Section \ref{subsect:phi.examp}
one can find examples of non-unital $\phi$ for which different
$\eta_g^{\riem}(\phi)$ are {\em not} identical.

\begin{theorem}\label{thm:inverse} We can rewrite the eigenvalue
problem (\ref{eigenvprob}) so that
\begin{eqnarray*}
\lambda_2^g(\phi,P)= \sup_{\alpha}
  \frac{\langle\widehat{\phi}(\alpha),(\Omega^g_P)^{-1} 
  [\widehat{\phi}(\alpha)] \rangle}
    {\langle \alpha,(\Omega^g_{\phi(P)})^{-1}[\alpha]\rangle}.
\end{eqnarray*}
where the supremum is now taken over  
$\left\{ \alpha \in \ran(\phi) :\tr [\Omega^g_{\phi(P)}]^{-1}(\alpha) = 0 
  \right\}$.
\end{theorem}

\proof
\begin{eqnarray*}
\lambda_2^g(\phi,P) & = & \sup_{A:\tr(A)=0}
    \frac{\langle\phi(A),\Omega^g_{\phi(P)}[\phi(A)]\rangle}
       {\langle A,\Omega^g_P[A]\rangle} \\
& = & \sup_{B:\tr [\Omega^g_P]^{-1/2}(B) = 0}
    \frac{\langle B [\Omega^g_P]^{-1/2} \circ \widehat{\phi} \circ
         \Omega^g_{\phi(P)} \circ \phi \circ [\Omega^g_P]^{-1/2}] B \rangle} 
              {\langle B, B \rangle} .
\end{eqnarray*} 
If we now write 
$\Gamma = [\Omega^g_{\phi(P)}]^{1/2} \circ \phi \circ [\Omega^g_P]^{-1/2}$,
we see that $\lambda_2^g(\phi,P)$ is the largest eigenvalue of
$\Gamma^* \Gamma$ where $\Gamma$ maps
\begin{eqnarray*}
\left\{ B : \tr [\Omega^g_P]^{-1/2}(B) = 0 \right\} \rightarrow
  \left\{ \beta \in \ran(\phi) :
    \tr [\Omega^g_{\phi(P)}]^{-1/2}(\beta) = 0 \right\}.
\end{eqnarray*} 
Since $\Gamma \Gamma^*$ and $\Gamma^* \Gamma$ have the same non-zero
eigenvalues, 
\begin{eqnarray*}
\lambda_2^g(\phi,P) & = &  
   \sup_{\beta:\tr [\Omega^g_{\phi(P)}]^{-1/2}(\beta) = 0}
    \frac{\langle \beta [\Omega^g_{\phi(P)}]^{1/2} \circ\phi \circ
   [\Omega^g_P]^{-1} \circ \widehat{\phi} \circ [\Omega^g_{\phi(P)}]^{1/2}] 
    \beta \rangle} {\langle \beta, \beta \rangle} \\
  & = &  \sup_{\alpha:\tr [\Omega^g_{\phi(P)}]^{-1}(\alpha) = 0}
   \frac{\langle \widehat{\phi}(\alpha)  
   [\Omega^g_P]^{-1} \widehat{\phi} (\alpha) \rangle} 
   {\langle \alpha,[\Omega^g_{\phi(P)}]^{-1} \alpha \rangle} .
\end{eqnarray*}

 If we apply this result with $\Omega_P^{\bur} = [ R_P + L_p]^{-1}$, 
it follows easily from the theorem above that
\begin{eqnarray}
\lambda_2^{\bur}(\phi,P) & = & \sup_{\alpha : \tr \phi(P) \alpha = 0}
    \frac { \tr \widehat{\phi}(\alpha) P \widehat{\phi}(\alpha)}
        { \tr \alpha \phi(P) \alpha}  .
\end{eqnarray}
It is tempting to write $\phi(P) \alpha = \beta = \phi(B)$ and replace
the constraint $\tr \phi(P) \alpha = 0$ by $\tr B = 0$.  The 
denominator would then become 
 $\langle \phi(B) [ \phi(P)]^{-1} \phi(B) \rangle$ which has the same
form as the numerator in (\ref{etadef.riem}) when $k(w) = \frac{w+1}{2w}$
(corresponding to $g = (w-1)^2$).  However, we there is no guarantee that
$\widehat{\phi}(\alpha) = \widehat{\phi}( [\phi(P)]^{-1}] B)$.
On the contrary, this cannot possibly hold because we would then
have that the $\lambda$ (and hence $\eta$) for the two extremal
functions $k(w) = \frac{2}{1+w}$ and  $k(w) = \frac{w+1}{2w}$
are inverses, which is inconsistent with 
$\lambda^g(\phi,P) \leq \eta^{\riem}_g(\phi) \leq 1$ 
(except in the case  $\lambda = 1$ which is not generic).
There is, however, a sense in which the operators associated
with these two extremal functions are inverses since
$ \Omega_P^{(w-1)^2} = R_P^{-1} + L_P^{-1} =  R_P^{-1}[R_P + L_P] L_P^{-1}
   =  R_P^{-1} [\Omega_P^{\bur}]^{-1} L_P^{-1}.$
It seems likely that if the $\eta_g^\riem$ for these two extremal functions
are equal, then all of them are.

Unlike the case of $\eta^{\RE}_g(\phi)$, we do expect that the
supremum for $\eta^{\riem}_g(\phi)$ is actually attained.
Indeed, we know that for each fixed $P$ the supremum in (\ref{lambdasup})
is attained for some $A \neq 0$ which satisfies the eigenvalue problem
(\ref{eigenvprob}).  As before, we can find a maximizing sequence
of density matrices $P_k$ for (\ref{eq:etariemsup}) so that
$\eta_g^{\riem}(\phi)= \lim_{k \rightarrow {\infty}} \lambda_2^g(\phi, P_k).$
For each $P_k$, let $A_k$ be the solution to the eigenvalue problem 
(\ref{eigenvprob}) for  $\lambda_2^g(\phi, P_k)$ normalized so that
$\tr |A_k| = 1.$  Then we can find a convergent subsequence for
which $P_k \rightarrow P \in {\cal D}$ and  $A_k \rightarrow A \neq 0$
since $\tr |A| = 1.$  It then follows that (\ref{eigenvprob})
holds for this $P, A$ with $ \lambda = \eta^{\riem}_g(\phi)$ (although
$P$ is only non-negative) which implies
\begin{eqnarray*}
\langle\phi(A),\Omega^g_{\phi(P)}[\phi(A)] \rangle =
  \eta^{\riem}_g(\phi) \langle A,\Omega^g_P(A)\rangle.
\end{eqnarray*}

\subsection{Bounds on contraction coefficients}

We first give an upper bound for $\eta_{\log}^{\riem}$ using 
\begin{eqnarray}
\eta^{\dob}(\phi) \equiv 
    \sup_{A \in T_* {\cal D}}\frac{\tr|\phi(A)|}{\tr|A|}.
\end{eqnarray}
This can be interpreted as the norm of $\phi$ regarded as an
operator on the Banach space of traceless matrices with norm $\tr|A|.$
Although the function $g(w) = |w-1|$ is {\em not} operator convex,
 $\eta^{\dob}(\phi)$ is analogous to the contraction coefficient
of the (non-differentiable) symmetric relative g-entropy 
$H_{|w-1|}(P,Q) = \tr |P-Q|$  which, however, 
 is {\em not} the relative g-entropy obtained
by using $g(w) = |w-1|$ in Definition \ref{relgentdef}.
Nevertheless, $\eta^{\dob}(\phi)$ is a natural and useful object
to consider.  It was shown in \cite{Rusk1} (see Theorem 2) that 
\begin{eqnarray}\label{eq:Dobrush}
\eta^{\dob}(\phi) =
  \half\sup\{\tr|\phi(E-F)|:~E,F ~ \hbox{1-dim projs}; EF=0 \}
\end{eqnarray}
where ``1-dim projs'' means that $E, F$ are one-dimensional projections in
$\dbar.$
The expression on the right in (\ref{eq:Dobrush}) shows that we
are justified in interpreting $\eta^{\dob}(\phi)$ as a non-commutative 
analogue of Dobrushin's coefficient of ergodicity.

\begin{theorem}  If $\phi$ is stochastic,
\begin{eqnarray}
\eta_{\log}^{\riem}(\phi) \leq \eta^{\dob}(\phi) \equiv 
  \sup_{A \in T_* {\cal D}}\frac{\tr|\phi(A)|}{\tr|A|} .
\end{eqnarray}
\end{theorem}
\proof 
The map $B\rightarrow(\Omega_P^{\log})^{-1}\circ
\widehat{\phi}\circ\Omega_{\phi(P)}^{\log}(B) \equiv \Phi_{\log}(B)$
 is positivity-preserving, as well as trace-preserving.  
The former follows from the integral
representations (\ref{eq:intrep.log.riem}) and (\ref{eq:omegaloginv})
for $\Omega_P^{\log}$ and its inverse together
with the fact that the composition of 
positivity-preserving maps is positive-preserving.  
Then taking the trace of
the absolute value of both sides of the eigenvalue problem
$\Phi[\phi(A)] = \lambda A$ and using
Theorem 1 of \cite{Rusk1} yields
\begin{eqnarray}
 \lambda \tr |A| = \tr |\Phi[\phi(A)]| \leq \tr |\phi(A)|.  \qed
\end{eqnarray}

Although we believe that this result holds for any $g$, we do not
have a proof except for the log. Our proof depended on the
observation that in the case of the log the map
$\Phi_g(B)  = (\Omega_P^g)^{-1}\circ
\widehat{\phi}\circ\Omega_{\phi(P)}^g(B)$ is 
positivity-preserving.  However, explicit examples can be found
to show that $\Phi_g$ is {\em not} positivity preserving in
general.  Indeed, although both $\Omega_P^{\bur}= [R_P + L_P]^{-1}$ and 
$\Omega_P^{(w-1)^2} = R_P^{-1} + L_P^{-1}$ are positive semi-definite
 with respect to the Hilbert-Schmidt inner product, they are not
positivity preserving in the sense of mapping positive operators
to positive operators.  The difference is analogous to the
difference between an ordinary matrix being positive semi-definite
and having positive elements.

We now consider lower bounds on $\eta_g^{\riem}(\phi).$
In \cite{Rusk1} it was shown that
\begin{eqnarray}\label{eta.lowbd1}
  \eta^{\dob}(\phi)  \leq  \sqrt{ \eta_{(w-1)^2}^{\riem}(\phi)}.
\end{eqnarray}
We now give
 a lower bound which holds for all $\eta_g^{\riem}(\phi)$ when
the map $\phi$ is unital, i.e., $\phi(I) = I.$
\begin{theorem}
 If $\phi$ is unital, 
\begin{eqnarray}\label{eta.lowbd2}
  \eta_g^{\riem}(\phi) \geq \sup_{\tr A = 0}
               \frac{\tr|\phi(A)|^2}{\tr|A|^2}. 
\end{eqnarray}
\end{theorem}
This is an immediate consequence of the definition (\ref{etadef.riem});
it also follows from Theorem \ref{thm:etariemsup} and the fact
that the right side of (\ref{eta.lowbd2}) is just $\lambda_2(\phi,I)$ 
when $\phi$ is unital.  The right side of (\ref{eta.lowbd2}) can also
be interpreted as the square of the norm
of $\phi$ regarded as an operator on the Banach space of traceless matrices
with Hilbert-Schmidt norm $\sqrt{\tr A^*A}.$
When $\phi$ is self-adjoint in the sense
$\widehat{\phi} = \phi$, every trace-preserving map is unital.

If $\phi$ maps ${\bf C}^{n \times n}$ to itself, then the results of this
section can be restated in terms of the eigenvalues and singular values
of $\phi$.  Since $\phi$ is trace-preserving, $\phi(B) = \Lambda B$
implies that either $\Lambda = 1$ or $\tr B = 0.$  If we restrict $\phi$
to the matrices with trace zero, then  $ \eta^{\dob}(\phi)$ is the
largest magnitude of an eigenvalue and for unital $\phi$
$\lambda_2(\phi,I)$ is the largest eigenvalue of $\widehat{\phi} \phi$.
Thus for unital stochastic maps,
$\lambda_2(\phi,I) = \Lambda_2(\widehat{\phi} \phi)$
where we have
continued our convention of using the subscript 2 for eigenvalues
of maps restricted to $T_*\dbar$.
If $\phi$
is self-adjoint, the two lower bounds (\ref{eta.lowbd1}) and 
(\ref{eta.lowbd2}) coincide and 
$\lambda_2(\phi,I) = \Lambda_2(\widehat{\phi} \phi) = [\Lambda_2(\phi)]^2$
in the usual sense of second largest eigenvalue of.  
For general  unital $\phi$, (\ref{eta.lowbd2}) is stronger since
\begin{eqnarray}
  \eta_{(w-1)^2}^{\riem}(\phi) \geq \lambda_2(\phi,I) = 
   \Lambda_2(\widehat{\phi} \phi) \geq  \left[ \eta^{\dob}(\phi) \right]^2.
\end{eqnarray}

We now explicitly state some conjectures which have already
been discussed.

\begin{conj} 
For each fixed $g \in {\cal G}$,
\begin{eqnarray}
\eta_g^\RE(\phi) = \eta_g^\riem(\phi)
   = \eta_g^\geo(\phi) \leq \eta_g^\dob(\phi) .
\end{eqnarray}
\end{conj}

\begin{conj} If $\phi$ is unital, then 
\begin{eqnarray}
\eta^{\riem}_g = \Lambda_2(\widehat{\phi} \phi)
 \equiv \sup_{\tr A = 0} \frac{\tr|\phi(A)|^2}{\tr|A|^2}
\end{eqnarray} for all $g \in {\cal G}.$
\end{conj}
If this conjecture holds, then for unital $\phi$ the contraction
coefficient $\eta^{\riem}_g$ is independent of $g$.  Theorem
(\ref{thm:non-unit}) at
the end of the next section contains an explicit example of a non-unital
stochastic map for which $\eta^{\riem}_g$ depends non-trivially on $g$;
therefore,  the hypothesis that $\phi$ be unital is essential.
In view of (\ref{eta.lowbd2}) it would suffice to show that
$\eta^{\riem}_g \leq \Lambda_2(\widehat{\phi} \phi)$

\subsection{Examples}\label{subsect:phi.examp}

We now consider some special classes of stochastic maps 
$\phi: {\cal A}_1 \rightarrow  {\cal A}_2.$  We begin by
looking at some maps for which
all contraction coefficients are easily seen to be zero or one.
We then consider maps from  ${\bf C}^{2 \times 2}$ to  ${\bf C}^{2 \times 2}$
which provide support for the conjecutres above.

We first consider the case in which ${\cal A}_2$ is one-dimensional,
e.g., $\phi$ projects onto a one-dimensional subalgebra (which
need not have an identity) of ${\cal A}_1.$
Then, since $\phi$ is trace-preserving and maps density
matrices to density matrices, we must have 
$\phi(P) = \phi(Q) ~\forall~ P,Q$ with $\tr \phi(P) = 1$ so that
$ \phi(P) \neq 0.$  Thus,
$H_g[\phi(P),\phi(Q)] = D_g[\phi(P),\phi(Q)] = 0 ~\forall~ P,Q$
which implies $\eta_g^\RE(\phi) = \eta_g^\geo(\phi) = 0.$
If $\tr B = 0$, then $\phi(B) = 0$.  (To see this note that one
can find $a,b$ such that $P = (aI + bB)$ is a density matrix.)  
Thus 
$\langle \phi(B) \Omega^g_{\phi(P)} \phi(B) \rangle = 0$
and $\tr |\phi(B)| = 0$ for all $B$ in $T_*\dbar$ which
implies $\eta_g^\riem(\phi) = \eta^{\dob}(\phi) = 0.$
We can summarize this as
\begin{theorem}\label{thm:etaeq0}
If the image of the stochastic map $\phi$ is one-dimensional,
then $\eta_g^\RE(\phi) = \eta_g^\riem(\phi) = \eta_g^\geo(\phi)
  = \eta^{\dob}(\phi) = 0$ for all $g \in {\cal G}.$
\end{theorem}

We next consider the important special case in which $\phi$ is a 
partial trace $\tau$.  In the simplest case, let
 $\tau: {\bf C}^{2n \times 2n} \rightarrow {\bf C}^{n \times n}$ 
be the map which takes
\begin{eqnarray}
{\bf M} =  \left( \begin{array} {cc}  A & B \\  C & D \end{array} \right)
  \rightarrow \tau({\bf M}) = A + D
\end{eqnarray}
where ${\bf M} \in {\bf C}^{2n \times 2n}$ has been written in
block form and $A, B, C, D \in {\bf C}^{n \times n}.$
Then the
homogeneity of relative entropy (see Definition \ref{relentdef}b)
 implies that for 
${\bf P} =  \left( \begin{array} {cc}  P & 0 \\  0 & P \end{array} \right)$
and 
${\bf Q} =  \left( \begin{array} {cc}  Q & 0 \\  0 & Q \end{array} \right)$
$$H_g({\bf P},{\bf Q}) = H_g(2P,2Q) = H_g(\tau({\bf P}),\tau({\bf Q}))$$ 
for any $g$, and similarly
$$\langle {\bf A},\Omega^g_{\bf P}({\bf A})\rangle =
    \langle 2A,\Omega^g_{2P}(2A)\rangle
  = \langle \tau({\bf A}),\Omega^g_{\tau({\bf P})}({\bf \tau({\bf A})})\rangle $$
 when
${\bf A} =  \left( \begin{array} {cc}  A & 0 \\ 0 & A \end{array} \right).$
From this, we easily see that  
\begin{eqnarray}\label{part.tr.eq}
\eta_g^\RE(\phi) = \eta_g^\riem(\phi) = \eta_g^\geo(\phi)
  = \eta^{\dob}(\phi) = 1,
\end{eqnarray}
where we have assumed implicitly that $\tau$ acts on the full algebra
of all $2n \times 2n$ matrices.
 
The partial trace described above is similar to a conditional expectation,
i.e., a map for which  ${\cal A}_2$ is a subalgebra (with identity)
of ${\cal A}_1$   and
$\phi(A) = A ~\forall~ A \in {\cal A}_2.$  Both
partial traces and conditional expectations are included in the
following
\begin{theorem}
If  the stochastic map $\phi$ is also an isomorphism
from a non-trivial subalgebra (with identity) of ${\cal A}_1$ to ${\cal A}_2$,
then $\eta_g^\RE(\phi) = \eta_g^\riem(\phi) = \eta_g^\geo(\phi)
  = \eta^{\dob}(\phi) = 1 $ for all $g \in {\cal G}.$
\end{theorem}

  Since every completely positive
map can be represented as a partial trace \cite{Lind}, this
might seem to suggest that $\eta = 1$ always holds.  However,
these representations involve multiple copies of the algebra,
so that the partial trace is not acting on the full algebra
in the higher dimensional space.   Thus, the representation
of ${\cal A}_1$ need necessarily not contain a subalgebra with
the desired isomorphism property.  Examples of maps
with $\eta < 1$ were already found in
\cite{Cetal} for commutative algebras, and two different
non-commutative  examples are given below.

We now state two results for maps
$\phi : {\bf C}^{2 \times 2} \rightarrow {\bf C}^{2 \times 2}.$
The proofs are postponed to a subsequent paper \cite{Rusk2}.  
Recall that any density matrix in 
${\bf C}^{2 \times 2}$ can be written in the form
$\half [I + \bw \dtsig]$ where $\bw \in {\bf R}^3$ and
$\sigma$ denote the vector of Pauli matrices.  The first
theorem provides evidence for the two conjectures at the
end of the previous section.

\begin{theorem}\label{thm:eta22} For the unital map $\phi_{\rmT}:
I + \bw \dtsig \raw I + \rmT \bw \dtsig $,
\begin{eqnarray*} 
  \eta_g^{\RE}(\phi_{\rmT}) =  \eta_g^{\riem}(\phi_{\rmT}) =
    \eta_g^{\geo}(\phi_{\rmT}) = \| \rmT \| ^2  ~~~\forall~  g \in {\cal G},
\end{eqnarray*} 
and $\eta^{\dob}(\phi_{\rmT})  = \| \rmT \|.$
\end{theorem}

The next example gives a non-unital stochastic map for which
$\eta_g^{\riem}(\phi)$ varies with $g.$
For $\alpha,\tau > 0$ with $\alpha + \tau \leq 1$, define
\begin{eqnarray}\label{counterex.phi}
\phi_{\alpha,\tau}[I + \bw \dtsig] = I + \alpha w_1 \sigma_1 + \tau \sigma_2 .
\end{eqnarray}
 It is easily seen to be stochastic because
the condition $\alpha + \tau \leq 1$ insures that it is a convex
combination of stochastic maps.
For $g_{s_0}(w) = (w-1)^2/(w+s_0)$ as in Example 3 of section \ref{sect:g.examp}
\begin{eqnarray*}\label{counterex.eta}
\eta_{g_{s_0}}^{\riem}(\phi) & = &  \sup_{0 \leq \omega \leq 1}
     \frac{ [(1-\tau^2 + (\rho -\alpha^2) \omega^2]~  [1 -  \omega^2]}
 {[1 - \tau^2 - \alpha^2 \omega^2]~
    [1 - \tau^2(1-\rho) - (1-\rho) \alpha^2 \omega^2 ] } \\        
 & \geq & \frac{\alpha^2}{1 - \left( \frac{1-s_0}{1+s_0} \right)^2 \tau^2 } \nonumber
\end{eqnarray*}
where $1-\rho = \frac{1-s_0}{1+s_0}$ and equality holds for $s_0 \approx 0.$
In particular, we can conlude
 
\begin{theorem}\label{thm:non-unit}
For the non-unital stochastic map $\phi$ given by (\ref{counterex.phi}),
there is an $S > 0$ such that for $s_0 \in [0,S)$, 
\begin{eqnarray*}
 \eta^{\riem}_{{s_0}}(\phi) = 
    \frac{\alpha^2}{1 - \left( \frac{1-s_0}{1+s_0} \right)^2 \tau^2}.
\end{eqnarray*}
Furthermore
\begin{eqnarray*}
  \eta^{\riem}_{(w-1)^2}(\phi) =  \frac{\alpha^2}{1 - \tau^2} < \alpha
   = \eta^{\dob}(\phi).
\end{eqnarray*}
\end{theorem}
If $s_1 \in (0,S)$, we have  $\eta^{\riem}_{{s_1}}(\phi) >  
\eta^{\riem}_{{s_0}}(\phi)  =  \frac{\alpha^2}{1 - \tau^2}.$

\bigskip

\noindent{\bf Acknowledgment:}  Portions of the this work were
done while the second author was an Affiliate of 
Department of Physics at Harvard University; she is grateful
to Professor Arthur Jaffe  for his hospitality.

\end{document}